\newcommand\lsim{\mathrel{\rlap{\lower4pt\hbox{\hskip1pt$\sim$}}
    \raise1pt\hbox{$<$}}}
\newcommand\gsim{\mathrel{\rlap{\lower4pt\hbox{\hskip1pt$\sim$}}
    \raise1pt\hbox{$>$}}}

\documentstyle[aas2pp4,epsf,float]{article}

\begin{document}

\title{Non-gaussian Features of Linear Cosmic String Models}
\author{P.\ P.\ Avelino}
\affil{Centro de Astrofisica, Universidade do Porto, Rua do Campo Alegre 823, 4150 Porto, Portugal}

\author{E.\ P.\ S.\ Shellard and J.\ H.\ P.\ Wu}
\affil{D.\ A.\ M.\ T.\ P., University of Cambridge,  Silver Street, Cambridge CB3 9EW, U.K.}

\and

\author{B. Allen}
\affil{Department of Physics, University of Wisconsin--Milwaukee
P.O. Box 413, Milwaukee, Wisconsin 53201, U.S.A.}


\begin{abstract}
We investigate the non-gaussian properties of cosmic-string-seeded
linear density perturbations with cold and hot dark matter backgrounds,
using high-resolution numerical simulations.
We compute the one-point probability density function
of the resulting density field,
its skewness, kurtosis, and genus curves for different smoothing 
scales. A semi-analytic model is then invoked to provide a 
physical interpretation of our results. We conclude that on scales 
smaller than $1.5{(\Omega h^2)}^{-1}$Mpc, perturbations seeded by
cosmic strings are very non-gaussian.
These scales may still be in a linear or mildly non-linear regime
in an open or $\Lambda$-universe with $\Gamma=\Omega h \lsim 0.2$.

\end{abstract}


\section{Introduction}
\label{intro}
At present,  the two main candidates for the origin of cosmic structure are
inflation and topological defects (for a review, see Vilenkin \& Shellard 1994).
Although both scenarios may produce a power spectrum of density
perturbations consistent with observations,
they have very different predictions 
regarding the statistical properties of the density field.
While most inflationary models produce gaussian random-phase
initial conditions, defect models produce non-gaussian perturbations
particularly on small scales.
New results from cosmic-string-seeded structure formation
using high-resolution simulations 
(\markcite{ASWAs,ASWAl}Avelino {\it et al.}~1997, 1998)
were encouraging for models with $\Gamma = \Omega h = 0.1$--$0.2$
(see also Battye {\it et al.}~1997);
both the mass fluctuation amplitude at $8 h^{-1}$Mpc, $\sigma_8$, and
the power spectrum shape of cosmic-string-induced
cold dark matter fluctuations,
${\cal P}(k)$,
were consistent within uncertainties
with
observational data
(Peacock \& Dodds 1994; Viana \& Liddle 1996).
However, because cosmic strings induce
non-gaussian density perturbations
on small scales,
the power spectrum alone is
insufficient to describe all the statistical properties of such a 
density field.
This is even more important in open or $\Lambda$-models because in 
those models the characteristic scales of the density field are shifted 
to larger scales relative to a flat model with $\Lambda=0$.

In this Letter we 
investigate the non-gaussian properties of the linear density field induced 
by cosmic strings using higher-order statistics such as the
skewness and the kurtosis of a one-point 
probability density function (PDF), as well as genus statistics.
The non-gaussian properties we reveal provide a significant observational 
signature for cosmic string-seeded structure formation models on small length-scales.
We note that previous analytic work has investigated the string-induced
velocity field on scales above several $h^{-1}$Mpc, which was inferred to be
gaussian (Vachaspati 1992; Moessner 1995), and that some of the features we 
study here were also observed in global topological defect models, notably for
textures (Park, Spergel, \& Turok 1991).
Past work on genus statistics in the context of topological defects was made
using toy models which incorporated some important features of the models
in question (Brandenberger, Kaplan \& Ramsey 1993;
Albrecht \& Robinson 1995; Avelino 1997).


Our first step in the present analysis 
was to perform high-resolution numerical simu\-lations of
cosmic string networks
in an expanding universe (Allen \& Shellard 1990) from which we subsequently computed 
the causally-sourced density field with either a cold or hot dark matter
(CDM or HDM) background.
The cosmic string simulations had
a dynamic range extending
from before the radiation-matter transition at $0.4 \eta_{\rm eq}$
through to deep into the matter era $8.4 \eta_{\rm eq}$, where $\eta_{\rm eq}$ 
is the conformal time at
radiation-matter density equality.
The structure formation simulation boxes contained $256^3$ grid-points and
their physical volume was in the range $(4$--$100 h^{-1}$Mpc$)^3$.
A much more detailed description of these 
methods is given by Avelino {\it et al.}~(1997, 1998).

\section{Non-gaussian test statistics}
\label{onepoint}
We first convolved the density field obtained above
with a top-hat window at different smoothing scales $R$,
and then calculated the resulting PDF $p(\nu)$,
where $\nu$ is the number of standard deviations from the mean.
The chosen smoothing scales were always at least three times larger than
the grid spacing to guarantee a sufficient resolution for the smoothing,
and
at least twelve times smaller than the box size to reduce sample variance
effects.
We then used the skewness and the kurtosis to test their gaussianity.
The skewness is defined as $M_3/M_2^{3/2}$,
where $M_i$ is the $i$-th central moment of a PDF.
Its sign indicates the skewed direction against
a normal distribution.
The kurtosis is defined as $M_4/M_2^2-3$ and
measures the size of the side tails against those of
a normal distribution.
Both statistics are zero for a gaussian distribution,
but will depart from zero when the distribution is non-gaussian.
We can therefore estimate a ``non-gaussian scale'' $R_{\rm ng}$,
below which both the skewness and the kurtosis depart significantly from zero.

Another useful statistic sensitive to the non-gaus\-sian features of
a density field is the genus,
which is employed to measure the topology of a continuous field.
Conceptually, the genus of a given surface can be defined as
$
g = {\rm \#\ of\  holes} - {\rm \#\ of\ isolated\ regions} +1.
$
A more useful  definition of genus is given by the Gauss-Bonnet theorem which 
relates the 
integrated Gaussian curvature (a local property) of a surface with the 
genus (a global property) of that surface:
\begin{equation}
\label{gabo}
\oint K dA =4 \pi (1-g),
\end {equation}
where $dA$ is a differential two-dimensional surface patch, and
$K=1/a_1 a_2$ is the Gaussian curvature of the patch with
$a_1$ and $a_2$ the two principal radii of curvature.
We used the numerical code developed by Avelino (1997) to calculate the
genus of these isodensity surfaces,
applying methods proposed by Gott, Mellot $\&$ Dickinson (1986).
Here we employed
the commonly used volume fraction parametrization of the genus curve,
that is, we define a new variable $\nu_f$ so that
\begin{equation}
  \label{vfraction}
  f={1 \over {{(2 \pi)}^{1/2}}} \int_{\nu_f}^\infty e^{-u^2/2} \, du,
\end{equation}
where $f$ is a previously calculated volume
fraction of the whole simulation box within which
the density is above a given threshold.
Having done this,
the genus curve becomes less-sensitive to the PDF 
which we studied separately.


\section{Results}
\label{results}

\begin{figure}[t]
  \plotone{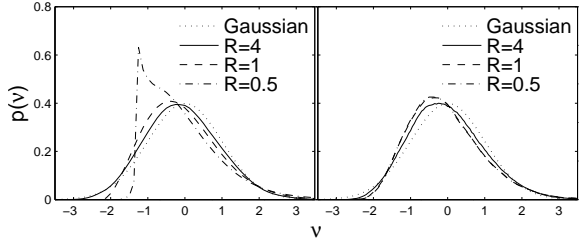}
  \caption{
  The one-point PDF $p(\nu)$ of the density field
  in the linear string$+$CDM (left) and string$+$HDM (right) models
  for
  different $R$ (in $(\Omega h^2)^{-1}$Mpc).
  Here $\sigma_{4\rm c}:\sigma_{1\rm c}:\sigma_{0.5\rm c}:
  \sigma_{4\rm h}:\sigma_{1\rm h}:\sigma_{0.5\rm h}
  \approx
  1:5.2:8.9:0.69:1.2:1.3$.
  All simulation curves are averaged from 6 realizations, with
  standard deviation less then $5\%$ of the amplitudes throughout.}
  \label{prob-d}
\end{figure}

\begin{figure}[t]
  \plotone{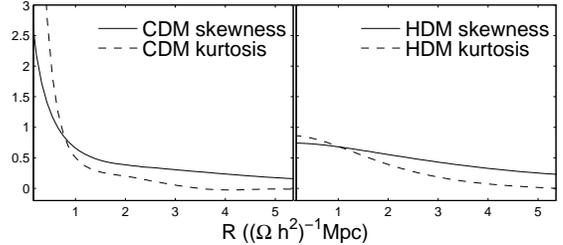}
  \caption{
  The skewness and kurtosis of $p(\nu)$ induced by cosmic strings.}
  \label{sk-ku}
\end{figure}

In Figure~\ref{prob-d} we plot
the one-point PDF $p(\nu)$ of
the density field in the linear string CDM and HDM models for
different smoothing scales $R$.
The skewness and kurtosis of $p(\nu)$
as functions of $R$ were plotted in Figure~\ref{sk-ku}.
Although on large scales string perturbations are almost indistinguishable 
from gaussian random-phase perturbations,
on scales smaller than $1.5 {(\Omega h^2)}^{-1}$Mpc 
the non-gaussian character is remarkably
distinct, especially in a CDM background.
\begin{figure}[t]
  \plotone{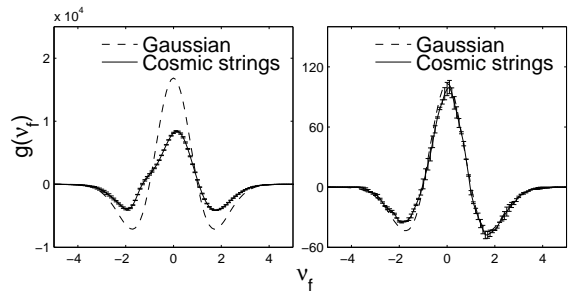}
  \caption{
  The genus curves
  in the CDM (left) and HDM(right) models.
  Here $R=0.4 (\Omega h^2)^{-1}$Mpc and
  the box size is $(50 h^{-1}{\rm Mpc})^3$.
  Error bars are taken from 6 realizations for each curve.}
  \label{gen}
\end{figure}

\begin{figure}[t]
\centering 
  \plotone{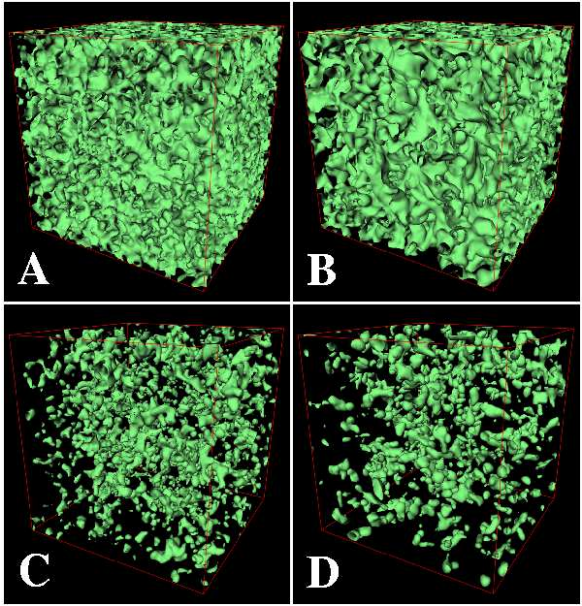}
  \caption{
  The isodensity surfaces with $\nu_f=0$ (A, B) and $\nu_f=2$ (C, D) for
  a gaussian field (A, C) and the string$+$CDM (B, D) model.    
  Here $R=0.4 (\Omega h^2)^{-1}$Mpc and
  the box size is $(25 h^{-1}{\rm Mpc})^3$.}
  \label{3D}
\end{figure}

In Figure~\ref{gen} we plot the genus curves $g(\nu_f)$ 
for
$R=0.4 (\Omega h^2)^{-1}$Mpc in both string$+$CDM and 
string$+$HDM models,
as well as those of gaussian 
random fields with correspondingly identical power spectra.
These curves have been smoothed using
three-point boxcar smoothing (see for example 
Vogeley {\it et al.} 1994) and were parametrized by the volume fraction
(\ref{vfraction}).
We can observe that on these small scales the genus curves of
the cosmic-string-seeded density fields deviate in a significant way 
from gaussian random fields.
As expected, the correlations among the phases of 
different Fourier modes imply a smaller number of independent regions
in real space and therefore
a smaller genus amplitude when compared with that of a gaussian random field.
This is illustrated in Figure~\ref{3D}, where two different isodensity thresholds for the 
CDM string model are directly compared with purely gaussian fluctuations
for $R= 0.4 (\Omega h^2)^{-1}$Mpc.  However, we have also verified that when 
smoothed on larger 
scales $R \gsim 3 (\Omega h^2)^{-1}$Mpc, cosmic-string and 
gaussian-random-phase genus curves are very close to each other in both
amplitude and shape. 
Finally, we note that the string$+$HDM model departure from a gaussian 
distribution is less apparent than for string$+$CDM model throughout all 
the analysis above.

\section{Discussion}
\label{disc_conc}

In order to interpret our results,
we invoked the semi-analytic model described in
Avelino {\it et al.}~(1997, 1998;
see also Albrecht \& Stebbins 1992)\markcite{ASWAl}.
This model was shown to be very accurate 
in reproducing the simulation power spectrum
in both matter-era and radiation-era scaling regimes, as well as during the 
radiation-matter-era transition. The power spectrum of string-induced density 
perturbations can be written as:
\begin{equation}
  \label{P_k}
  {\cal P}(k)=16\pi^2 G^2 \mu^2 \int_{\eta_{\rm i}}^{\eta_{\rm f}}
  |{\widetilde{\cal G}}(k;\eta_0,\eta)|^2
  {\cal F}(k, \eta) d\eta\,,
\end{equation}
where $\mu$ is the string energy density per unit length,
${\widetilde{\cal G}}(k;\eta_0,\eta)$ the Fourier transform of the 
appropriate Green's functions,
and ${\cal F}(k, \eta)$ the structure function.
At a given time,
${\cal F}(k, \eta)$ has a turn-over scale at $k_{\xi}\approx20/\eta$
which reflects the correlation length of cosmic strings 
$\xi \approx d_{\rm H}/3$.
At a particular time, 
the perturbations induced on scales larger than the string correlation length 
are generated by many string elements and, therefore, they are nearly gaussian 
according to the central limit theorem.
On the other hand, perturbations induced on smaller scales are very non-gaussian 
because they can be either very large within the string-induced wakes 
or else very small in regions  
outside these wakes. In consequence, we can roughly divide the power spectrum 
of cosmic-string-seeded density perturbations 
into two parts:
a nearly gaussian part generated when the string correlation length
was smaller than the scale under consideration,
and a strongly skewed non-gaussian part generated 
when the string correlation length was larger.  We shall abuse strict 
definitions and call these simply the `gaussian' and `non-gaussian' contributions
respectively.  In terms of the structure function in (\ref{P_k}), we 
can make the split ${\cal F}_{\rm g}(k,\eta) = {\cal F}(k,\eta)$ for 
$k<k_\xi$ (${\cal F}_{\rm g}=0$ for $k>k_\xi$) and, similarly, 
${\cal F}_{\rm ng}(k,\eta)= {\cal F}(k,\eta)$ for $k>k_\xi$.  
${\cal F}_{\rm g}$ 
corresponds to the left-hand side of the structure function from its peak to
the causal compensation cut-off where ${\cal F}_{\rm g}\propto k^4$ for 
$k <\!<k_\xi$.
${\cal F}_{\rm ng}$ can be largely identified with the 
imprint of string wakes and behaves as ${\cal F}_{\rm ng}\propto k^{-2}$ for
$k>k_\xi$.
\begin{figure}[t]
  \plotone{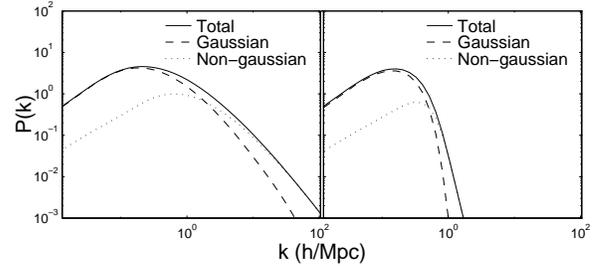}
  \caption{
  The estimated gaussian and non-gaussian contributions as a
  comparison with the total power spectrum
  in the CDM (left) and HDM (right) models.}
  \label{pow-g-ng}
\end{figure}

\begin{figure}[t]
  \plotone{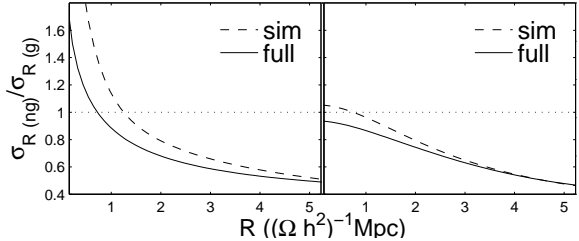}
  \caption{
  The ratio
  $\sigma_{R{\rm (ng)}}/\sigma_{R{\rm (g)}}$
  against $R$ for both CDM (left) and HDM (right) models.
  The solid lines have a full dynamic range
  going from $\eta=0$ to today,
  while the dashed lines have the same dynamic range of our simulations.}
  \label{sigma}
\end{figure}

Integrating (\ref{P_k}) with this gaussian/non-gaussian split, we can compare
the relative contributions to the total power spectrum in CDM and HDM models,
as illustrated in Figure~\ref{pow-g-ng}. 
This plot demonstrates that these two contributions become 
comparable at $k \sim 2 h{\rm Mpc}^{-1}$ for the CDM model
and $k \sim 0.5h{\rm Mpc}^{-1}$ for the HDM model,
taking a Hubble parameter $H_0=70 {\rm km\ s^{-1}Mpc^{-1}}$.
We further calculated the variance of the density fields from these
non-gaussian and gaussian contributions when smoothed with a top-hat window 
of radius $R$:
\begin{equation}
  \label{sigma_R}
  \sigma_R^2
  =
  4\pi \int_0^\infty |{\widetilde w}(kR)|^2 {\cal P}(k) k^2dk \,,
\end{equation}
where ${\widetilde w}(x)=3j_{\scriptscriptstyle 1}(x)/x$ is the
Fourier transform of a top-hat window and
$j_{\scriptscriptstyle 1}(x)$ the spherical Bessel function of order one.
Figure~\ref{sigma} plots
the ratio between the standard deviations
of the non-gaussian and gaussian density perturbations,
$\sigma_{R{\rm (ng)}}/\sigma_{R{\rm (g)}}$,
against $R$
for both CDM and HDM models.
The criterion $\sigma_{R{\rm (ng)}}/\sigma_{R{\rm (g)}}\gsim 1$,
gives an estimate for the non-gaussian scale $R_{\rm ng}$.
In Figure~\ref{sigma},
we also consider the effect of the limited dynamic range of our simulations,
by comparing with the full semi-analytic integral (\ref{P_k}).  The simulations
slightly enhance non-gaussianity essentially because there is more 
`gaussian' power missing from early times $\eta < 0.4\eta_{\rm eq}$.  Nevertheless, 
we have checked that the power lost
$(\sigma_{\rm (full)}-\sigma_{\rm (sim)})/\sigma_{\rm (full)}$
on scales $R \geq 1 (\Omega h^2)^{-1}$Mpc is always
$\leq 10 \%$ for the non-gaussian contribution,
$\leq 35 \%$ for the gaussian contribution, and
$\leq 25 \%$ for the total power.
This confirms the reliability of our non-gaussian analysis based on these
simulations.

The consistency between Figure~\ref{sk-ku} and Figure~\ref{sigma}
should be evident, notably for the skewness. The physical reason for the
much stronger non-gaussian features
on small scales can now be interpreted as the following.
Before the radiation-matter equality $\eta_{\rm eq}$, the growth of density 
perturbations was suppressed.
This means that even on small scales
the contribution to the power spectrum from the matter era is very important.
Adding the fact that
by the time of $\eta_{\rm eq}$ the string correlation length
had already grown to several $(\Omega h^2)^{-1}$Mpc,
the perturbations generated during the matter era on scales smaller than this
will add an important non-gaussian contribution to
the total power spectrum.
This also explains the peak in the PDF on small scales
(see Figure~\ref{prob-d}).
The limited dynamic range of our simulations tends to
produce a sharp peak on the PDF for small smoothing scales
because of the presence of voids, that is, 
regions left largely unperturbed by the 
transition era strings.  However,
even for the full dynamic range a similar peak, though slightly smoother, 
should still emerge.
This is the origin of the non-gaussianity on smaller scales, the existence 
of which is not affected by 
the limited dynamic range as we demonstrate in Figure~\ref{sigma}.
The reason why the HDM model is more gaussian than the CDM model is that
the neutrino free-streaming length prevents small-scale perturbations from 
emerging until late times.  This significantly reduces the HDM 
power at large $k$, precisely 
where the non-gaussian contribution is dominant in
the CDM model (see Figure~\ref{pow-g-ng}). 
This also accounts for the smaller genus amplitude for the HDM model (see Figure~\ref{gen}).

We also note that our results were obtained in the linear regime.
When the density field evolves into the non-linear regime at later times,
the skewness increases as a direct result of the constraint
that the density contrast $\delta \geq -1$.
However, when compared with observations,
the skewness of our linear cosmic-string-seeded result
for any reasonable value of $\Gamma$ is well below
that of the QDOT survey (Saunders {\it et al.}~1991). Our results 
are also consistent with those of Canavezes {\it et al.} (1997) as 
we find that cosmic-string and gaussian-random-phase genus curves 
are very similar in the scale range where no strong phase correlations 
were found by the genus analysis of the PSCz survey.
For gaussian models, it is well known that in the non-linear regime
the skewness is equal to its initial value plus
a contribution due to the non-linear collapse.
This contribution scales proportionally to the rms density fluctuation
$\sigma$ (Fry \& Scherrer 1994).
Similarly, the non-linear contribution to the kurtosis
also has a simple scaling dependence on $\sigma$,
depending on whether or not the initial density field is gaussian
(Chodorowski \& Bouchet 1995).
This indicates that non-linear clustering preserves the non-gaussianity
inherent in the linear evolution,
but further investigation using $N$-body simulations will be required
to establish if the non-gausian signature of cosmic string models
is potentially measurable.

\section{Conclusion}

We conclude that on length scales 
smaller than\ $1.5 {(\Omega h^2)}^{-1}{\rm Mpc}$ perturbations seeded by
cos\-mic strings are very non-gaussian,
especially in the context of a CDM model.
In an open or $\Lambda$-universe with 
$\Gamma=\Omega h \sim 0.15$, this scale will be shifted to $10 h^{-1}$Mpc,
which may still be in the linear or mildly non-linear regime, thus
potentially providing a strong empirical test for cosmic string models.
It has been suggested that such non-gaussianity may imply that it is difficult
in string models to deduce the parameter $\beta=\Omega_0^{0.6}/b$ from 
observations of density and velocity fields (van de Bruck, 1997).  
However, our results indicate otherwise on large scales because
we find that string perturbations are very similar to
gaussian-random phase fluctuations, when smoothed on sufficiently large scales.

\acknowledgements

We thank Carlos Martins and
Robert Caldwell for useful conversations.
P.~P.~A.\ is funded by 
JNICT (PRAXIS XXI/BPD/9901/96). 
E.~P.~S.~S.\ acknowledges support from PPARC grant GR/L 21488.
J.~H.~P.~W.\ is funded by CVCP (ORS/96009158) and
by the Cambridge Overseas Trust (UK).
B.~A.\ acknowledges support from NSF grant PHY95-07740.
This work was performed on COSMOS, the Origin2000 owned by the UK
Computational Cosmology Consortium, supported by Silicon Graphics/Cray
Research, HEFCE and PPARC.




\end{document}